\documentclass[12pt]{article}
\usepackage{graphicx}
\begin{document}
 \noindent {\footnotesize\it Astronomy Letters, 2010 Vol. 36, No. 7, pp.~506-–513}

 \noindent
 \begin{tabular}{llllllllllllllllllllllllllllllllllllllllllllll}
 & & & & & & & & & & & & & & & & & & & & & & & & & & & & & & & & & & & & &  \\\hline\hline
 \end{tabular}

 \vskip 1.5cm
 \centerline {\large\bf Determining the Orientation Parameters of the}
 \centerline {\large\bf ICRS/UCAC2 System Using the Kharkov Catalog}
 \centerline {\large\bf of Absolute Stellar Proper Motions}
 \bigskip
 \centerline {V.V. Bobylev$^1$, P.N. Fedorov$^2$, A.T. Bajkova$^1$, and V.S. Akhmetov$^2$}
 \bigskip
 \centerline {\small\it
 $^1$Pulkovo Astronomical Observatory, Russian Academy of Sciences, St-Petersburg}
 \centerline {\small\it $^2$Karazin Kharkov National University, ul. Sumskaya 35, Kharkov, 61022 Ukraine}
 \bigskip
 \bigskip

{\bf Abstract--}The absolute proper motions of about 275 million
stars from the Kharkov XPM catalog have been obtained by comparing
their positions in the 2MASS and USNO--A2.0 catalogs with an epoch
difference of about 45 yr for northern-hemisphere stars and about
17 yr for southern-hemisphere stars. The zero point of the system
of absolute proper motions has been determined using 1.45 million
galaxies. The equatorial components of the residual rotation
vector of the ICRS/UCAC2 coordinate system relative to the system
of extragalactic sources have been determined by comparing the XPM
and UCAC2 stellar proper motions:
$\omega_{x,y,z}=(-0.06,0.17,-0.84)\pm(0.15,0.14,0.14)$ mas
yr$^{-1}$. These parameters have been calculated using about 1
million faintest UCAC2 stars with magnitudes $R_{UCAC2}>16^m$ and
$J>14^m.7$,for which the color and magnitude equation effects are
negligible.



\section*{INTRODUCTION}

Since 1998, the International Celestial Reference System (ICRS)
has been realized by the International Celestial Reference Frame
(ICRS), which is represented by the catalog of positions of
quasars and other extragalactic radio sources. This catalog was
extended to the optical range by the HIPPARCOS and Tycho Catalogue
(1997).

Kovalevsky et al. (1997) established that the ICRS/HIPPARCOS
coordinate system had no residual rotation relative to an inertial
coordinate system with an error of $\pm0.25$ mas yr$^{-1}$ along
three axes.

Subsequently, as new data became available, Bobylev (2004b) found
the following rotational parameters of the HIPPARCOS system
relative to the extragalactic coordinate system:
 $\omega_x=0.04\pm0.15$ mas yr$^{-1}$,
 $\omega_y=0.18\pm0.12$ mas yr$^{-1}$, and
 $\omega_z=-0.35\pm0.09$ mas yr$^{-1}$
(the rotation components in the equatorial coordinate system).
This result is confirmed by a kinematic analysis of HIPPARCOS
stars (Bobylev 2004a) and hundreds of thousands of faint Tycho-2
and UCAC2 stars (Bobylev and Khovritchev 2006). On the whole, it
is also consistent with the results by Boboltz et al. (2007), who
analyzed the positions and proper motions of 46 radio stars and
obtained the mutual orientation parameters of the optical
realization (HIPPARCOS) and the radio system:
 $e_x=-0.4\pm2.6$ mas,
 $e_y= 0.1\pm2.6$ mas, and
 $e_z=-3.2\pm2.9$ mas, as well as the components of the residual
rotation vector:
 $\omega_x= 0.55\pm0.34$ mas yr$^{-1}$,
 $\omega_y= 0.02\pm0.36$ mas yr$^{-1}$, and
 $\omega_z=-0.41\pm0.37$ mas yr$^{-1}$.
Boboltz et al. (2007) reached the right conclusion that there are
no significant rotations within the error limits. Note also the
estimate of the angular velocity of rotation of the HIPPARCOS
system relative the DE403 and DE405 systems of ephemerides
obtained by Chernetenko (2008) by analyzing the observations of
asteroids:
 $|\omega| =0.94±0.20$ mas yr$^{-1}$,
 where the components of the vector found are
 $\omega_x= 0.12\pm0.08$ mas yr$^{-1}$,
 $\omega_y= 0.66\pm0.09$ mas yr$^{-1}$, and
 $\omega_z=-0.56\pm0.16$ mas yr$^{-1}$.
 This result suggests that either the
dynamical DE403 and DE405 models need to be improved or the
HIPPARCOS system needs to be corrected. As we see, one of the
components, namely, $\omega_z$, differs significantly from zero in
several cases. Determining this parameter by an independent method
is a very topical problem.

Based on data from the 2MASS (Skrutskie et al. 2006) and
USNO--A2.0 (Monet 1998) catalogs, Fedorov et al. (2009) derived
the absolute proper motions of about 275 million stars fainter
than $12^m$ at the Karazin Kharkov National University. For this
catalog, we use the abbreviation XPM. The XPM stars cover the
entire celestial sphere with the exception of a small region near
the Galactic-center direction (Fig. 1b). The stellar proper
motions were obtained by comparing the stellar positions in the
2MASS Point Source Catalog and USNO--A2.0 with an epoch difference
of about 45 yr for northern-hemisphere stars and about 17 yr for
southern-hemisphere stars. The zero point of the system of
absolute proper motions (absolutization corrections) was
determined using about 1.45 million galaxies from the 2MASS
catalog. Thus, the XPM catalog is an independent realization of an
inertial coordinate system. The most significant systematic zonal
errors in USNO--A2.0 were eliminated before the derivation of
proper motions. The mean formal absolutization error is less than
1 mas yr$^{-1}$, while the random error in the proper motion is
3--8 mas yr$^{-1}$, depending on the magnitude. The current
version of the XPM catalog contains stellar positions in ICRS for
the epoch J2000, original absolute stellar proper motions, and B,
R, J, H,and K magnitudes.

In this paper, our task is to determine the residual rotation
vector of the optical realization of the ICRS/HIPPARCOS system
relative to the coordinate system specified by extragalactic
sources.

As a realization of the ICRS/HIRRARCOS system, we use the UCAC2
catalog (Zacharias et al. 2004), which extends the system to stars
as faint as $R_{UCAC2}\approx16^m.5$. The XPM catalog of proper
motions acts as a realization of an inertial coordinate system.
The task is accomplished by comparing the proper motions of common
stars in the XPM and UCAC2 catalogs based on which the mutual
rotation parameters (the components of vector $\omega$) are
determined.

\section{THE METHOD}

To determine $\omega(\omega_x,\omega_y,\omega_z)$, we use the well
known equations (Lindgren and Kovalevsky 1995)
$$
\displaylines{\hfill
 \Delta\mu_\alpha\cos\delta=
 \omega_x\cos\alpha\sin\delta   
 +\omega_y\sin\alpha\sin\delta-
 \omega_z\cos\delta,\hfill\llap(1)\cr\hfill
 \Delta\mu_\delta=-\omega_x\sin\alpha+ \omega_y\cos\alpha,
 \hfill\llap(2)
 }
$$
where the XPM--UCAC2 stellar proper motion differences are on the
left-hand sides. The system of conditional equations (1) and (2)
is solved by the least-squares method.

We identified a total of about 36 million common stars in the XPM
and UCAC2 catalogs; 63 XPM fields with unreliable absolutization
were excluded--these are located along the Galactic equator, in
the zone of avoidance, where there are very few or no galaxies.

Figure 1 shows an example of the distribution of XPM stars over
the celestial sphere. About 1.6 million stars with $B_{USNO-B}$
magnitudes in the range $17^m.00-17^m.02$ were used to construct
the diagram.

In Fig. 1a, we clearly see a nonuniformity of the distribution due
to a strong concentration of stars to the Galactic plane.
Obviously, if the system of equations (1) and (2) is solved using
individual stars, then the solution will be biased, because the
Galactic equator zone, in which an overwhelming majority of stars
are located, will have the greatest weight. To get rid of this
nonuniformity, we apply Charlier's well-known method using
equal-area fields. The essence of the method is that despite the
difference in the number of stars, a unit weight is assigned to
each field when the system of conditional equations (1) and (2) is
solved. We divided the sky into 432 fields. Because of various
gaps, only about 380 such fields are actually used. The gaps
mostly stem from the fact that the UCAC2 catalog contains no data
in the northern--sky zone at $\delta>54^\circ$.

Comparison of the XPM and UCAC2 stars showed that the mean
dispersions of the stellar proper motion differences do not exceed
9 mas yr$^{-1}$ in both coordinates. In this case, the estimate of
$9/\sqrt 2=6.4$ mas yr$^{-1}$ is in good agreement both with the
declared accuracy of the UCAC2 proper motions and with the
estimates of the accuracy of the XPM proper motions.

To eliminate erroneous identifications, we used the following
constraints on the stellar proper motion differences:
 $|\Delta\mu_\alpha\cos\delta|<50$ mas yr$^{-1}$ and
 $|\Delta\mu_\delta|<50$ mas yr$^{-1}$.

\section{RESULTS}
\subsection{Analysis of the Differences}

In Fig. 2, the mean XPM.UCAC2 proper motion differences are
plotted against the $J$ magnitude; in Fig. 3, the components of
the vector $\omega(\omega_x,\omega_y,\omega_z)$ found from the
differences of the same stars are plotted against the $J$
magnitude. About 30 million stars were used to construct Figs. 2
and 3.

We see from Fig. 3 that the components $\omega_x$, and $\omega_y$
do not differ significantly from zero in the entire range of
magnitudes under consideration. In contrast, $\omega_z$ depends
strongly on the magnitude. Note that $\omega_z$ is determined only
from Eq. (1) and, hence, its determination is influenced only by
differences of the form $\mu_\alpha\cos\delta$.

It follows from Fig. 2 that the magnitude dependence of the
$\omega_z$ differences in the range $12^m<J<15^m$ can be fitted by
a linear trend with a coefficient of the magnitude equation (ME)
$-0.6\pm0.05$ mas yr$^{-1}$ per magnitude. This trend can be
associated with the presence of ME in the UCAC2 catalog. Indeed,
Bobylev and Khovrichev (2006) showed that the proper motions of
UCAC2 stars with magnitudes in the range $R_{UCAC2}=12^m-15^m$ are
distorted by ME in $\mu_\alpha\cos\delta$ with a coefficient of
$-0.6\pm 0.05$ mas yr$^{-1}$ per magnitude. The magnitude
dependence of the PUL3SE--UCAC2 differences in
$\mu_\alpha\cos\delta$ is similar in pattern to that in Fig. 2.

Since Bobylev and Khovrichev (2006) found no noticeable magnitude
dependence of the PUL3SE--UCAC2 differences $\mu_\delta$, it can
be suggested that no ME of this type is present in the XPM
catalog. It follows from Fig. 2 that in the range $10^m<J<13^m$,
the ME coefficient in the differences $\mu_\delta$ is $-0.6$ mas
yr$^{-1}$ per magnitude.

Studying ME is the subject of a separate investigation, while to
accomplish our task, it will suffice to take the faintest UCAC2
stars without any noticeable ME. Stars with $R_{UCAC2}>16^m$ are
quite suitable for this purpose. As follows from Fig. 2, bright
stars should also be removed to minimize the influence of ME in
$\mu_\delta$ of the XPM catalog.

In Fig. 4, the components of the vector .found are plotted against
the J magnitude for a sample of stars with $R_{UCAC2}>16^m$. Since
there is a constraint on RUCAC2, we observe an increase in color
index, for example, $R_{UCAC2}-J$, with decreasing $J$. As can be
seen from the figure, two components, $\omega_y$ and $\omega_z$,
behave stably, while $\omega_x$ exhibits a noticeable trend and
tends to $-1$ mas yr$^{-1}$ for $J<14^m$. This behavior of
$\omega_x$ is related to the presence of a color equation in one
of the catalogs being analyzed, because for $J<14^m$ the color
index, on average, $R_{UCAC2}-J>2^m$, i.e., the stars have a
significant reddening. The percentage of such stars is low, but it
is better to exclude them from consideration.

The data in Fig. 4 suggest that the influence of both magnitude
and color equations is small for the proper motion differences of
stars with $J>14^m$ and $R_{UCAC2}>16^m$.

To study the influence of the color equation on the color equation
is present only in differences of the form quantities being
determined, Eqs. (1) and (2) were solved separately. The sample
and the approach are the same as those used in constructing Fig.
4, but the magnitude range is wider, $J>13^m.0$. The results are
reflected in Fig. 5, from which we see that a significant color
equations is present only in differences of the form
$\Delta\mu_\alpha\cos\delta$ and affects mainly the determination
of the parameter $\omega_x$. It is easy to see that for the
dependence of $\omega_x$ in the range $R-J=1m.5–3^m$, the
coefficient of the linear trend is $\approx-2.8$ mas yr$^{-1}$ per
magnitude.

Using the differences of 3 146 504 stars with $J>14^m.0$ and
$R_{UCAC2}>16^m.0$, we found the following parameters by solving
the system of equations (1) and (2):
$$
\displaylines{\hfill
 \omega_x=-0.08\pm0.14~\hbox {mas yr$^{-1}$},\hfill\llap(3)\cr\hfill
 \omega_y=+0.21\pm0.14~\hbox {mas yr$^{-1}$},\hfill\cr\hfill
 \omega_z=-0.95\pm0.13~\hbox {mas yr$^{-1}$}.\hfill\cr
 }
$$
For this sample, $\overline{R}_{UCAC2}=16^m.2$,
$\overline{J}=14^m.8$,and the mean color index
$\overline{R-J}=1^m.6$.

\subsection{The Kinematics of Sample Stars}

To study the properties of our sample of stars, we found the
kinematic parameters of the linear Ogorondikov-Milne model. The
method is described in detail in Bobylev and Khovrichev (2006).
The conditional equations can be written as
$$\displaylines{\hfill
\mu_{l}\cos b=
       X_{\odot}\sin l-Y_{\odot}\cos l-
\hfill\llap(4) \cr\hfill
   -M_{\scriptscriptstyle32}^{\scriptscriptstyle-}\cos l\sin b
   -M_{\scriptscriptstyle13}^{\scriptscriptstyle-}\sin l\sin b
   +M_{\scriptscriptstyle21}^{\scriptscriptstyle-}\cos b
   +M_{\scriptscriptstyle12}^{\scriptscriptstyle+}\cos 2l\cos b
   -M_{\scriptscriptstyle13}^{\scriptscriptstyle+}\sin l\sin b+
\hfill\cr\hfill
   +M_{\scriptscriptstyle23}^{\scriptscriptstyle+}\cos l\sin b
  -0.5(M_{\scriptscriptstyle11}^{\scriptscriptstyle+}
  -M_{\scriptscriptstyle22}^{\scriptscriptstyle+})\sin 2l\cos b,
\hfill \cr\hfill
 \mu_b=
    X_{\odot}\cos l\sin b+Y_{\odot}\sin l\sin b-Z_{\odot}\cos b
\hfill\llap(5) \cr\hfill
   +M_{\scriptscriptstyle32}^{\scriptscriptstyle-}\sin l
   -M_{\scriptscriptstyle13}^{\scriptscriptstyle-}\cos l
-0.5M_{\scriptscriptstyle12}^{\scriptscriptstyle+}\sin 2l\sin 2b
   +M_{\scriptscriptstyle13}^{\scriptscriptstyle+}\cos l\cos 2b+
\hfill\cr\hfill
   +M_{\scriptscriptstyle23}^{\scriptscriptstyle+}\sin l\cos 2b
-0.5(M_{\scriptscriptstyle11}^{\scriptscriptstyle+}
    -M_{\scriptscriptstyle22}^{\scriptscriptstyle+})\cos^2 l\sin 2b
+0.5(M_{\scriptscriptstyle33}^{\scriptscriptstyle+}
    -M_{\scriptscriptstyle22}^{\scriptscriptstyle+})\sin 2b.
\hfill }
$$In this writing (without allowance for the individual distances),
all of the sought-for unknowns are expressed in mas yr$^{-1}$. The
Galactic proper motion components averaged in each Charlier field
are on the left-hand sides; $l$ and $b$ are the Galactic
coordinates. $X\odot, Y\odot,$ and $Z\odot$ are the peculiar solar
velocity components
 $M_{\scriptscriptstyle12}^{\scriptscriptstyle-}$,
 $M_{\scriptscriptstyle13}^{\scriptscriptstyle-}$, and
 $M_{\scriptscriptstyle23}^{\scriptscriptstyle-}$ are the components
of the solid-body rotation vector of a small solar neighborhood
around the Galactic $Z$, $Y$ and $X$ axes, respectively. The
quantity $M_{\scriptscriptstyle12}^{\scriptscriptstyle-}$ (mas
yr$^{-1}$) is related to the Oort constant $B$ (km s$^{-1}$
kpc$^{-1}$) via the proportionality coefficient 4.74.

Each of the quantities
 $M_{\scriptscriptstyle12}^{\scriptscriptstyle+}$,
 $M_{\scriptscriptstyle13}^{\scriptscriptstyle+}$, and
 $M_{\scriptscriptstyle23}^{\scriptscriptstyle+}$ describes the deformation
in the corresponding plane. The quantity
$M_{\scriptscriptstyle12}^{\scriptscriptstyle+}$ (mas yr$^{-1}$)
is related to the Oort constant $A$ (km s$^{-1}$ kpc$^{-1}$) via
the proportionality coefficient 4.74. The diagonal components of
the local deformation tensor
 $M_{\scriptscriptstyle11}^{\scriptscriptstyle+}$,
 $M_{\scriptscriptstyle22}^{\scriptscriptstyle+}$, and
 $M_{\scriptscriptstyle33}^{\scriptscriptstyle+}$
describe the overall contraction or expansion of the entire
stellar system. If only the stellar proper motions are used, then
one of the diagonal terms of the local deformation tensor is known
to remain uncertain. Therefore, we determine differences of the
form
$(M_{\scriptscriptstyle11}^{\scriptscriptstyle+}-
 M_{\scriptscriptstyle22}^{\scriptscriptstyle+})$ and
$(M_{\scriptscriptstyle33}^{\scriptscriptstyle+}-
 M_{\scriptscriptstyle22}^{\scriptscriptstyle+})$.

To estimate the mean distance to the sample stars, we use a
statistical method. As the known peculiar solar velocity relative
to the local standard of rest, we take the values from Dehnen and
Binney (1998):
$(U_\odot,V_\odot,W_\odot)=(10.00,5.25,7.17)\pm(0.36,0.62,0.38)$
km s$^{-1}$. We calculate the parallax using two formulas:
$\pi_U=4.74\cdot X_\odot/U_\odot$ and $\pi_W=4.74\cdot
Z_\odot/W_\odot$, where $X_\odot$ and $Z_\odot$ are the stellar
group velocity components found, expressed in mas yr$^{-1}$. Since
the component $Y_\odot$ is noticeably distorted by the asymmetric
drift velocity (Dehnen and Binney 1998), this projection is not
used to determine the group parallaxes. We find the distance $d$
from the relation $d=1/\pi$. The table gives two estimates
obtained from the velocities $U_\odot$ and $W_\odot$.

The results of solving the system of equations (4) and (5) by the
least-squares method are presented in the table. To eliminate the
stars with large proper motions that can spoil the statistics, we
used a constraint on the magnitude of the tangential stellar
velocity, $\sqrt{(\mu_\alpha\cos\delta)^2+(\mu_\delta)^2}<150$ mas
yr$^{-1}$. The solutions were obtained for the same sample of
stars from which solution (3) was found, but with two different
sets of stellar proper motions--from the XPM and UCAC2 catalogs.

As follows from the table, all kinematic parameters of the
Ogorodnikov-Milne model are determined slightly more accurately
from the UCAC2 proper motions. The solutions obtained from UCAC2
are in excellent agreement with their analysis performed by
Bobylev and Khovrichev (2006) based on different samples from this
catalog. There are slight differences for a number of parameters
found using the XPM catalog. These include the solar velocity
components $X_\odot$ and $Z_\odot$ and, as a result, the
difference in the estimates of the statistical distance $d_U$ and
$d_W$. There are also differences in the estimates of the Oort
constants $A$ and $B$. In compiling our sample of stars for
solution (3), we took faint UCAC2 stars, while for the XPM catalog
these stars ($\overline{B}=16^m.6$)are by no means faint (the
limiting magnitude is $B\approx21^m$). The above differences in
these parameters are probably attributable to the presence of
small (we got rid of the large ones) magnitude and color equation
effects in the XPM catalog, which require their careful study and
elimination to obtain reliable data in analyzing the Galactic
kinematics.

For the goals of this paper, we are most interested in the values
of the parameter
 $M_{13}^{-}= 0.57\pm0.10$ mas yr$^{-1}$ from XPM data and
 $M_{13}^{-}=-0.39\pm0.09$ mas yr$^{-1}$ from UCAC2 data.
The magnitude of their XPM–UCAC2 differences is 0.96 mas
yr$^{-1}$, in agreement with the $\omega_z$ magnitude of solution
(3). This is because in the Galactic coordinate system, the
direction of the celestial pole is close to the direction of the
Galactic $Y$ axis. Therefore, the rotation around the equatorial
$Z$ axis manifests itself in the Galactic coordinate system mainly
as the rotation around the Galactic $Y$ axis that, in our case, is
described by the parameter $M_{13}^{-}$.

\subsection{The Vector $\omega$.}

Using the proposed constraints, we got rid of the significant
manifestations of the magnitude and color equations. However, as
we see from Figs. 4 and 5, even for $J>14^m$ there are small
trends for all of the quantities being determined. Therefore, for
the final solution, we used the differences of 1 145 768 faintest
stars with $J>14^m.7$ and $R_{UCAC2}>16^m.0$, which, in our
opinion, deserve the greatest confidence. In Figs. 4 and 5, two
rightmost points correspond to this magnitude range. For this
sample, ${\overline J}=14.^m9$ and the mean color index
${\overline {R-J}}=1.^m4$. As a result, we found the parameters
$$
\displaylines{\hfill
 \omega_x=-0.06\pm0.15~\hbox {mas yr$^{-1}$},\hfill\llap(4)\cr\hfill
 \omega_y=+0.17\pm0.14~\hbox {mas yr$^{-1}$},\hfill\cr\hfill
 \omega_z=-0.84\pm0.14~\hbox {mas yr$^{-1}$},\hfill\cr
 }
$$
which are the main result of this work.

\section{DISCUSSION}

The values of the components $\omega_x$ and $\omega_y$ we found do
not differ significantly from zero. The value of
$\omega_z=-0.84\pm0.14$ mas yr$^{-1}$ differs significantly from
zero. Our results are qualitatively in good agreement with those
of several most extensive individual programs used to reference
the HIPPARCOS catalog to the system of extragalactic sources
(Kovalevsky et al. 1997). For example, the Kiev program (Kislyuk
et al. 1997) yielded
 $\omega_x=-0.27\pm0.80$ mas yr$^{-1}$,
 $\omega_y=+0.15\pm0.60$ mas yr$^{-1}$, and
 $\omega_z=-1.07\pm0.80$ mas yr$^{-1}$.
Note also the results of the Pulkovo program (Bobylev et al.
2004):
 $\omega_x=-0.98\pm0.47$ mas yr$^{-1}$,
 $\omega_y=-0.03\pm0.38$ mas yr$^{-1}$, and
 $\omega_z=-1.66\pm0.42$ mas yr$^{-1}$.
Our parameters are also in good agreement with the present-day
results of the analysis of a long-term series of asteroid
observations (Chernetenko 2008) noted in the Introduction.

The manifestations of the magnitude and color equations found in
the bright part of the current XPM catalog are attributable to the
peculiarities of deriving its proper motions, because the stellar
images in the 2MASS and USNO--A2.0 catalogs were obtained in the
near-infrared and optical ranges, respectively.

Using a sufficiently large number of comparison stars in solving
(4) allowed the sought-for parameters to be determined with a high
accuracy. Our constraints give hope that the results obtained are
not distorted by the magnitude and color equation effects.

\subsection{CONCLUSIONS}

Comparison of the intermediate version of the XPM catalog and
UCAC2 showed that the mean dispersions of the stellar proper
motion differences are $\approx9$ mas yr$^{-1}$ in both
coordinates. This gives an estimate for the mean random error of
the stellar proper motions from the catalogs in external
convergence $\approx6$ mas yr$^{-1}$, which is in good agreement
both with the declared accuracy of UCAC2 and with the estimates of
their accuracy in XPM.

We established that the XPM--UCAC2 proper motion differences, both
$\Delta\mu_\alpha\cos\delta$ and $\Delta\mu_\delta$, have
significant nonlinear magnitude dependences.

We believe that the magnitude equation in
$\Delta\mu_\alpha\cos\delta$ and $\Delta\mu_\delta$ is related to
its presence in the UCAC2 and XPM catalogs, respectively. In the
range $10^m<J<13^m$,the coefficient of the linear trend for the
magnitude equation in $\Delta\mu_\delta$ is $\approx-0.6$ mas
yr$^{-1}$ per magnitude.

We showed that a significant dependence on color, the coefficient
of the linear trend of which in the range $R_{UCAC2}–J=1^m.5-3^m$
for $\omega_x$ is $\approx-2.8$ mas yr$^{-1}$ per magnitude, is
present in $\mu_\alpha\cos\delta$. To minimize these effects, we
proposed a number of constraints on the magnitude and color of
stars. We showed that acceptable parameters
$\omega_x,\omega_y,\omega_z$ could be obtained for $J>14^m.0$ and
$R_{UCAC2}>16^m$ (solution (3)). Our kinematic control based on
the linear Ogorodnikov-Milne model indicated that the sample of
XPM stars used has no significant kinematic deviations.

The most reliable components of the residual rotation vector of
the ICRS/UCAC2 coordinate system relative to the system of
extragalactic sources,
 $(\omega_x,\omega_y,\omega_z)=(-0.06,0.17,-0.84)\pm(0.15,0.14,0.14)$ mas yr$^{-1}$,
were calculated using 1 145 768 faintest UCAC2 stars with
 $R_{UCAC2}>16^m$ and $J>14^m.7$.

The components $\omega_x,\omega_y$, and $\omega_z$ found can be
used to derive the most probable parameters of referencing the
optical realization of the ICRS/HIPPARCOS system to the system of
extragalactic sources.

Of great interest is a further kinematic analysis of the absolute
proper motions for faint XPM stars. However, the magnitude and,
particularly, color equation effects in this catalog should be
carefully studied and removed.

\subsection*{ACKNOWLEDGMENTS}

We are grateful to the referees for useful remarks that
contributed to an improvement of this paper. The Russian authors
are grateful to G.T. Bajkova for help in working with the data.
This study was supported by the Russian Foundation for Basic
Research (project no. 09--02--90443--Ukr\_f) from the Russian side
and by the Contest for joint basic research projects
``DFFD--RFFD--2009'' (project no. F28.2/042) from the Ukrainian
side.

\bigskip
{\bf REFERENCES}

{\small

 D.A. Boboltz, A.L. Fey, W.K. Puatua, et al., Astron. J. 133, 906 (2007).

 V.V. Bobylev, Pis'ma Astron. Zh. 30, 289 (2004a) [Astron. Lett. 30, 251 (2004)].

 V.V. Bobylev, Pis'ma Astron. Zh. 30, 930 (2004b) [Astron. Lett. 30, 785 (2004)].

 V.V. Bobylev and M.Yu. Khovrichev, Pis'ma Astron. Zh. 32, 676 (2006) [Astron. Lett. 32, 608 (2006)].

 V.V. Bobylev, N.M. Bronnikova, and N.A. Shakht, Pis'ma Astron. Zh. 30, 519 (2004) [Astron. Lett. 30, 469 (2004)].

 Yu. A. Chernetenko, Pis'ma Astron. Zh. 34, 296 (2008) [Astron. Lett. 34, 266 (2008)].

 W. Dehnen and J.J. Binney, Mon. Not. R. Astron. Soc. 298, 387 (1998).

 P.N. Fedorov, A.A. Myznikov, and V.S. Akhmetov, Mon. Not. R. Astron. Soc. 393, 133 (2009).

 The HIPPARCOS and Tycho Catalogues, ESA SP-1200 (1997).

 V.S. Kislyuk, S.P. Rybka, A.I. Yatsenko, et al., Astron. Astrophys. 321, 660 (1997).

 J. Kovalevsky, L. Lindegren, M.A.C. Perryman, et al., Astron. Astrophys. 323, 620 (1997).

 L. Lindegren and J. Kovalevsky, Astron. Astrophys. 304, 189 (1995).

 D.G. Monet, Bull. Am. Astron. Soc. 30, 1427 (1998).

 M.F. Skrutskie, R.M. Cutri, R. Stiening, et al., Astron. J. 131, 1163 (2006).

 N. Zacharias, S.E. Urban, M.I. Zacharias, et al., Astron. J. 127, 3043 (2004).


}

\newpage
{
\begin{table}[p]                                                
\caption[]{\small\baselineskip=1.0ex\protect
 Kinematic parameters of the Ogorodnikov–Milne model
 }
\begin{center}
\begin{tabular}{|c|r|r|c|}\hline
   Parameters   &  XPM   &  UCAC2    \\\hline

      $X_\odot$,~mas/yr&  $ 1.53\pm0.11$ & $ 2.55\pm0.09$ \\
      $Y_\odot$,~mas/yr&  $ 6.16\pm0.10$ & $ 7.59\pm0.09$ \\
      $Z_\odot$,~mas/yr&  $ 1.14\pm0.11$ & $ 2.28\pm0.09$ \\

   $M_{21}^{+}$,~mas/yr&  $ 1.89\pm0.14$ & $ 2.72\pm0.12$ \\
   $M_{32}^{-}$,~mas/yr&  $-0.28\pm0.11$ & $-0.10\pm0.09$ \\
   $M_{13}^{-}$,~mas/yr&  $ 0.57\pm0.10$ & $-0.39\pm0.09$ \\
   $M_{21}^{-}$,~mas/yr&  $-2.12\pm0.11$ & $-2.39\pm0.09$ \\

$M_{11-22}^{+}$,~mas/yr&  $-0.19\pm0.27$ & $-0.21\pm0.23$ \\
   $M_{13}^{+}$,~mas/yr&  $-0.16\pm0.13$ & $ 0.10\pm0.11$ \\
   $M_{23}^{+}$,~mas/yr&  $ 0.24\pm0.13$ & $-0.06\pm0.11$ \\
$M_{33-22}^{+}$,~mas/yr&  $-0.07\pm0.28$ & $ 0.12\pm0.24$ \\
 \hline
    $d_U$, kpc &  $ 1.4\pm0.4$ & $ 0.8\pm0.2$ \\
    $d_W$, kpc &  $ 1.3\pm0.4$ & $ 0.7\pm0.2$ \\

  $A$, km/s/kpc &  $  8.9\pm0.6$ & $ 12.9\pm0.5$ \\
  $B$, km/s/kpc &  $-10.1\pm0.5$ & $-11.3\pm0.4$ \\
 \hline
\end{tabular}
\end{center}
\vskip60mm
\end{table}
}

\newpage

\newpage
\begin{figure}[p]
{
\begin{center}
 \includegraphics[width=100mm]{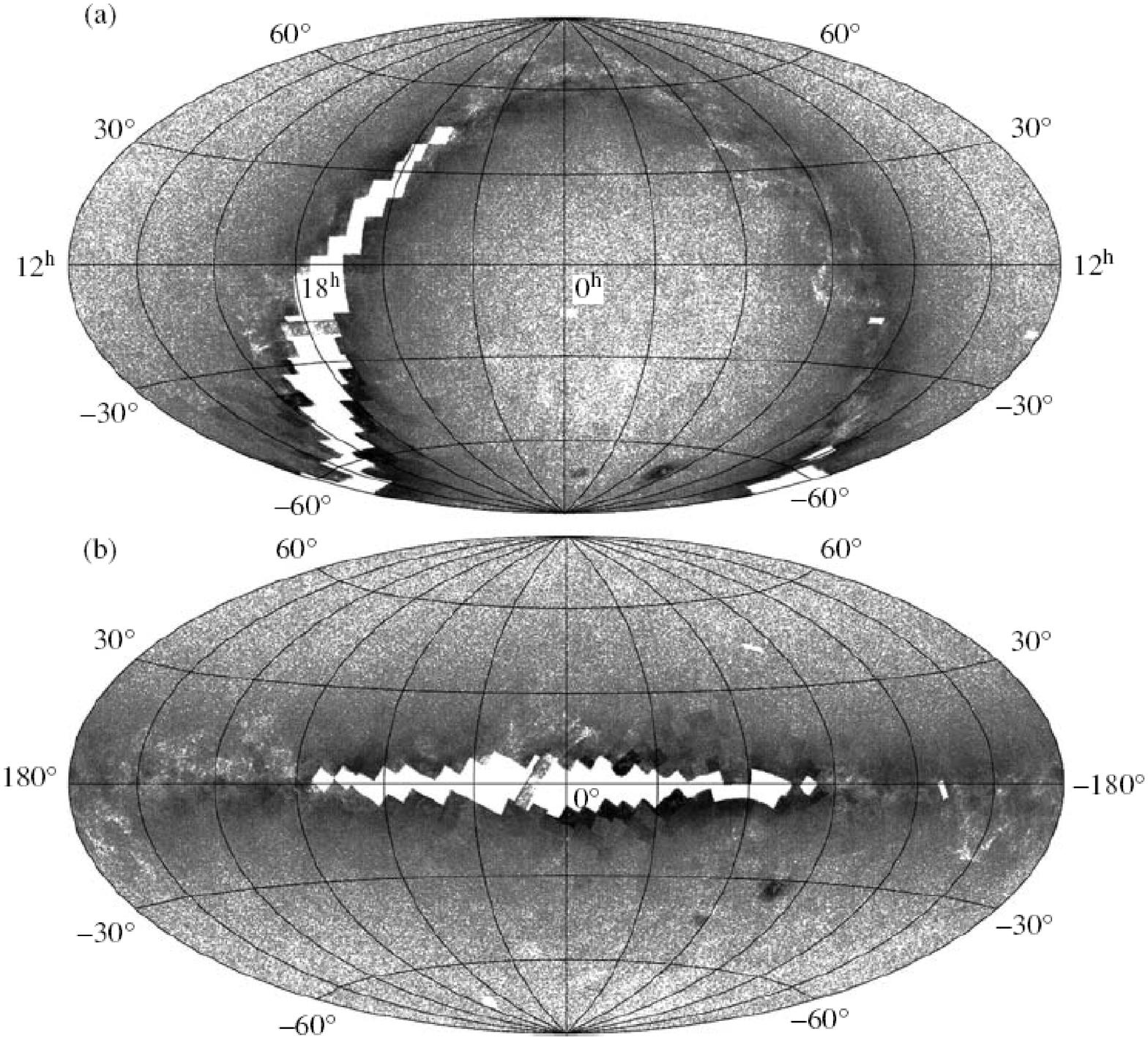}
\end{center}
} Fig.~1. Distributions of the sample of XPM $B=17^m$ stars over
the celestial sphere in the equatorial (a) and Galactic (b)
coordinate systems. The empty regions mark 63 fields with
unreliable absolutization.
\end{figure}

\newpage
\begin{figure}[p]
{
\begin{center}
 \includegraphics[width=100mm]{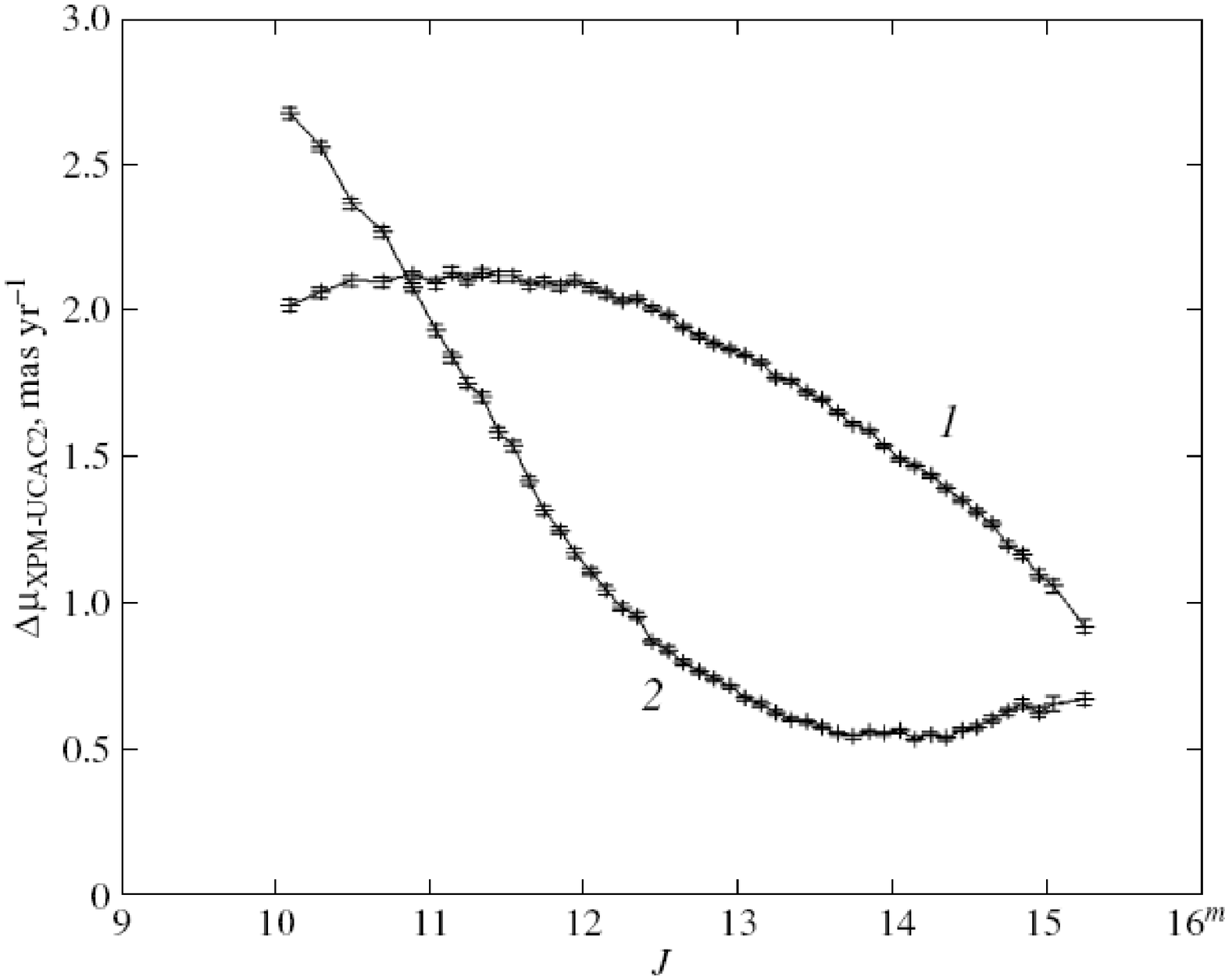}
\end{center}
} Fig.~2. Mean XPM--UCAC2 stellar proper motion differences
$\mu_\alpha\cos\delta$ (1) and $\mu_\delta$ (2) versus $J$
magnitude.
\end{figure}

\newpage
\begin{figure}[p]
{
\begin{center}
 \includegraphics[width=100mm]{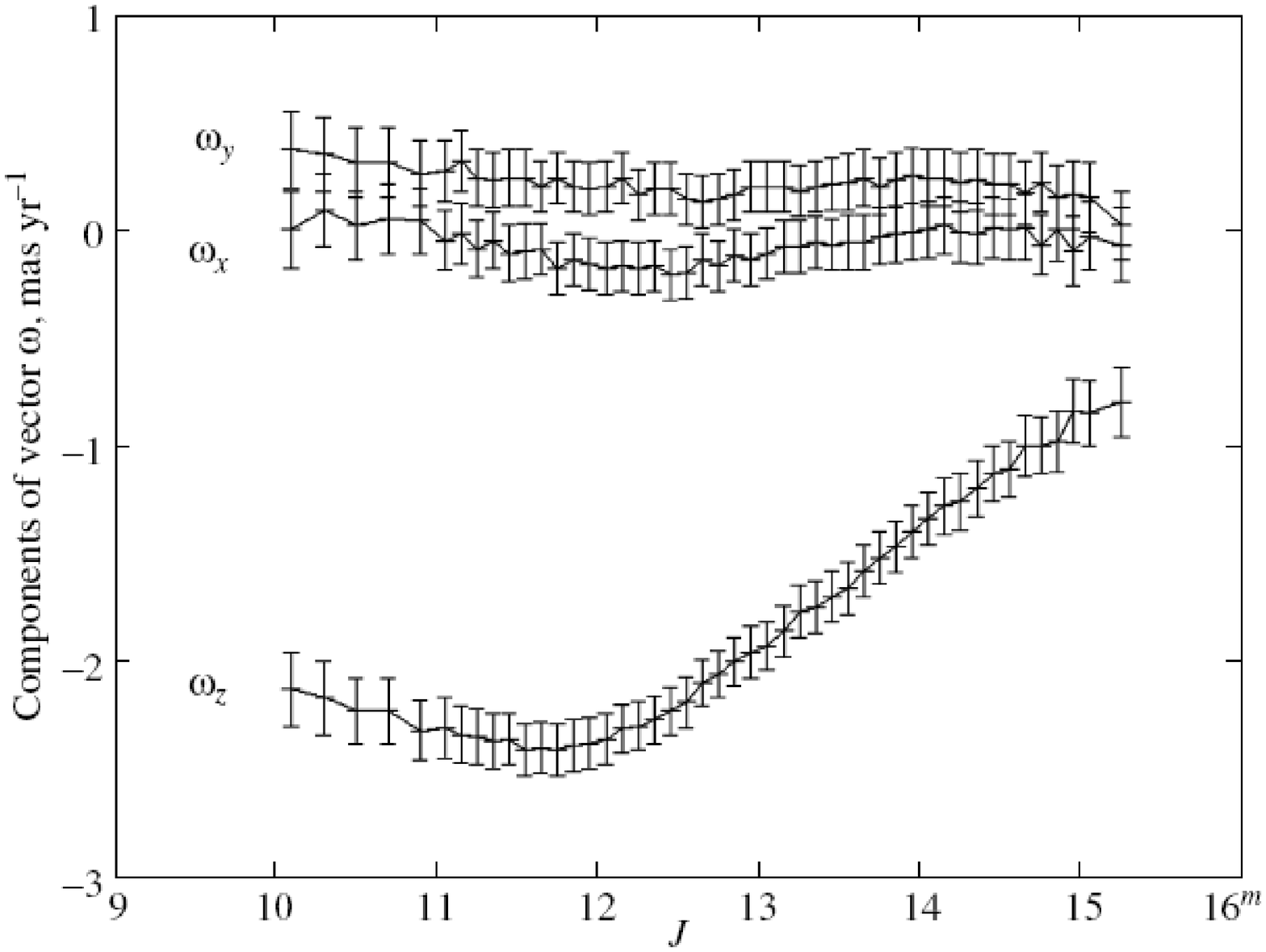}
\end{center}
} Fig.~3. Components of the residual rotation vector of the
ICRS/UCAC2 coordinate system relative to the system of
extragalactic objects found by comparing the XPM and UCAC2 stars
versus $J$ magnitude.
\end{figure}

\newpage
\begin{figure}[p]
{
\begin{center}
 \includegraphics[width=100mm]{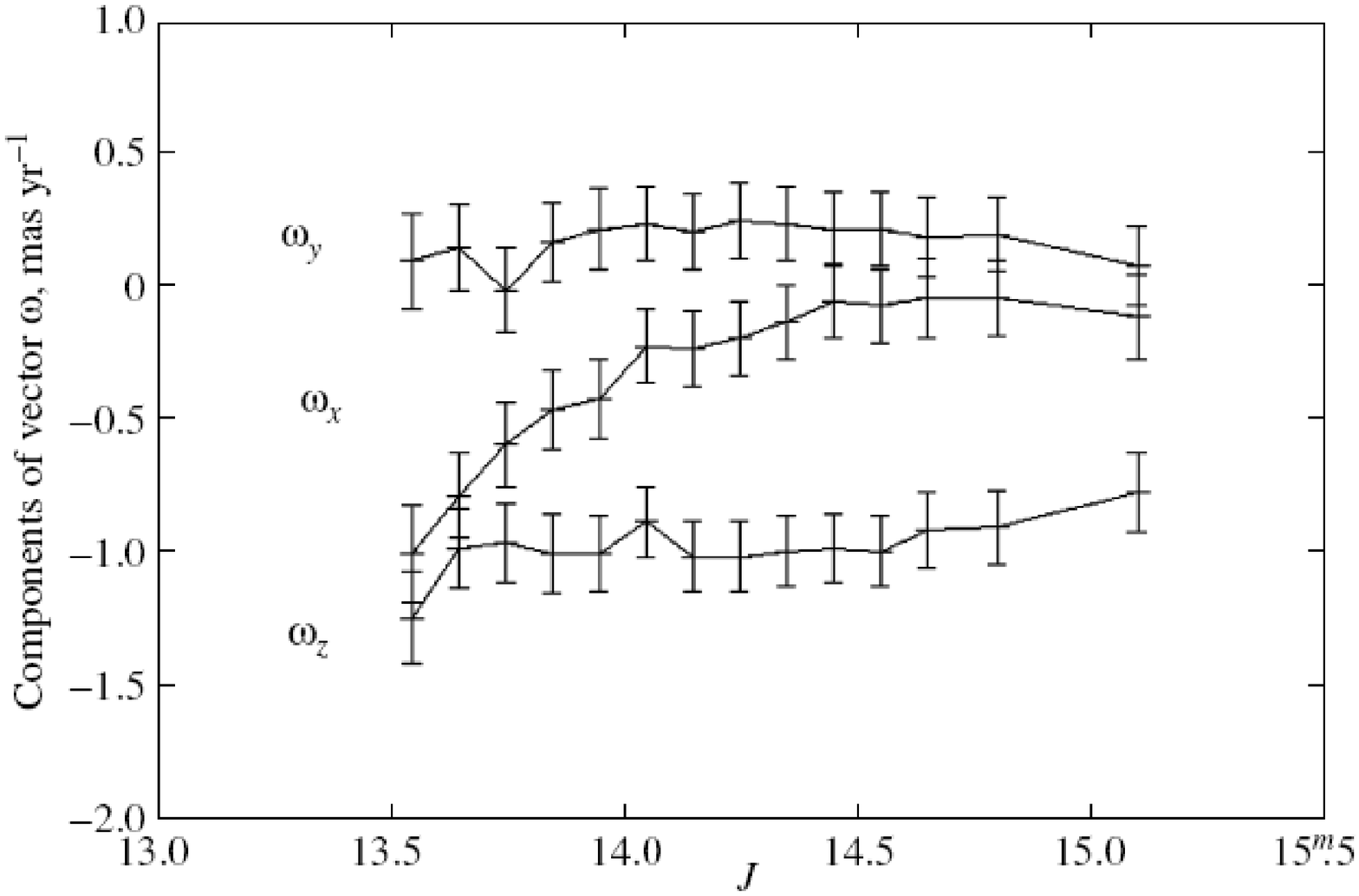}
\end{center}
} Fig.~4. Components of the residual rotation vector of the
ICRS/UCAC2 coordinate system relative to the system of
extragalactic objects found by comparing the XPM and UCAC2 stars
versus $J$ magnitude for a sample of stars with $R_{UCAC2}>16^m$.
\end{figure}

\newpage
\begin{figure}[p]
{
\begin{center}
 \includegraphics[width=100mm]{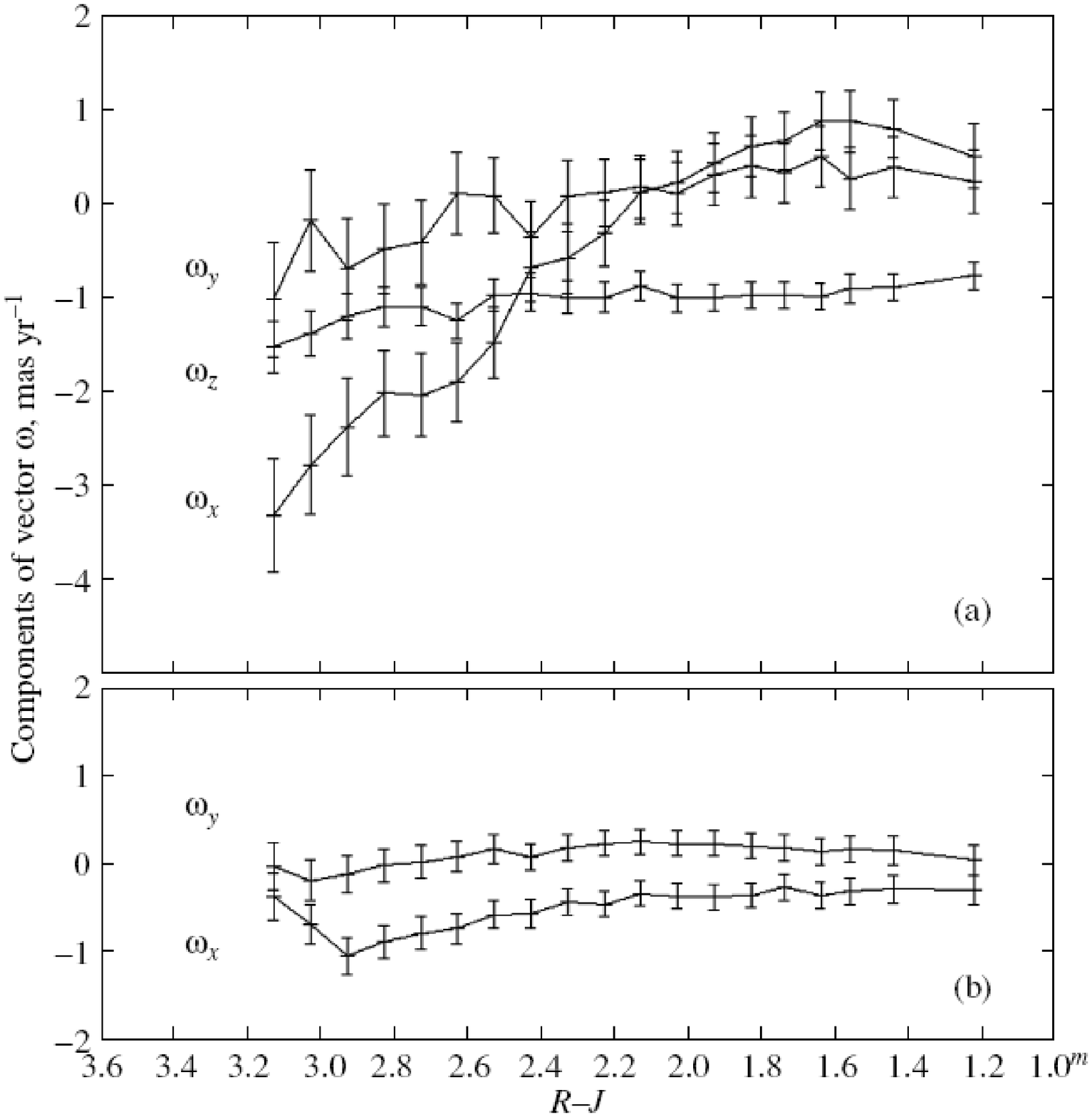}
\end{center}
} Fig.~5. Components of the vector $\omega$ versus $R_{UCAC2}-J$
color index found by separately solving Eqs. (1) and (2): (a) only
from Eq. (1) and (b) only from Eq. (2).
\end{figure}

\end{document}